\documentclass[twocolumn,showpacs,preprintnumbers,amsmath,amssymb,pre]{revtex4}

\usepackage{graphics}
\usepackage{graphicx}
\usepackage{amsmath,amssymb,epsfig,picinpar,graphics,float,amssymb,verbatim}
\newcommand{\eq}[1]{Eq.\ \ref{#1}}
\newcommand{\sech}{\text{sech}}
\newcommand{\sn}{\text{sn}}
\begin{document}

\title{Solitons on H-bonds in proteins}
\author{Francesco d'Ovidio}
\thanks{dovidio@fysik.dtu.dk}
\affiliation{Center for Quantum Protein, Dept. of
Physics, Technical University of Denmark, DK 2800 Lyngby,
Denmark}
\author{Henrik Georg Bohr}
\thanks{hbohr@fysik.dtu.dk}
\affiliation{Center for Quantum Protein, Dept. of
Physics, Technical University of Denmark, DK 2800 Lyngby,
Denmark}
\author{Per-Anker Lindg{\aa}rd}\thanks{p.a.lindgard@risoe.dk}
\affiliation{Dept. of Mat. Research, Ris{\o} National Laboratory, 4000 Roskilde, Denmark}
\date{\today}

\begin{abstract}
A model for soliton dynamics on a hydrogen-bond network
in helical proteins is proposed. It employs in three dimensions the formalism of
fully integrable Toda lattices which admits
phonons as well as solitons along the hydrogen-bonds of the helices.
A simulation of the three dimensional Toda lattice system shows that
the solitons are spontaneously created and are stable and moving along the helix axis. A
perturbation on one of the three H-bond lines forms solitons on
the other H-bonds as well.
The robust solitary wave may explain very
long-lived modes in the frequency range
of 100 cm$^{-1}$ which are found in recent X-ray laser experiments.
The dynamics parameters of the Toda lattice are in
accordance with the usual Lennard-Jones parameters used for realistic
H-bond potentials in proteins.
\end{abstract}
\pacs{05.45-a,87.10.+e}
\maketitle

\section{Introduction}

The present paper is addressing important new experimental protein
results that have come out, and in particular 
the recent InfraRed (IR)  measurements of
long-lived excitations at 118 cm$^{-1}$ using the
pump probe technique on Bacteriorhodopsine \cite{xie02}. 
These results are interesting because excitations at 
these energies do not correspond to any
local vibrational mode.
Since they are in the far infrared region they have been interpreted as
\emph{collective} modes, that is, modes that involve a large number of
amino acids, possibly involving large scale deformations of the protein.
If so, one would expect strong damping and short life times because of steric
hindrance from the remaining protein and from the surrounding solvent. 
The relevance of these observations lies in the fact
that such states, corresponding to large protein domains,
provide information on the dynamics and stability of secondary and higher
structures, and thus on the functions and the conformational changes of a
protein.
However, the phenomenon of collective modes is not fully
understood from a theoretical point of view, since both models and
numerical simulations face the difficulty of a large number of degrees
of freedom with complex interactions. In order to shed some light on the paradoxical long life times 
of the modes an alternative interpretation was given in terms of hydrogen-bond excitations
running along the $\alpha$-helix without causing major large scale deformations
\cite{lindgaard02}. The response at $\sim$ 100 cm$^{-1}$, typically
found in poly-amides, are generally accepted to be due to phonons
extended over H-bonds chains \cite{colaianni95}.
Poly-amides form molecular chains consisting of hydrogen bonded units of
(H-N-C=O) which are similar to those found running almost parallel to the axis 
along the $\alpha$-helices, and connecting every third residue. 
Every residue is additionally connected to its neighbors along the spiral
spine by a strong peptide bond. Bacteriorhodopsine consists of
7 connected $\alpha$-helices, with an average length of 25 residues. 
To test that the simple chain picture is relevant for
Bacteriorhodopsine, we here perform numerical
simulations using the special 3D-architecture of the $\alpha$-helix.
We find that apart from 
phonon modes, there are long-lived excitations which may
serve as energy reservoir for other excitations. 
They are not directly observable as a finite frequency response signal
in the excitation spectrum. They 
are localized modes, or solitons, traveling along the hydrogen
strands of an $\alpha-$helix and coupled with the peptide bonds. Their
long life time is due to the fact that such waves, being
non-oscillatory and localized,
interact only weakly with the other modes of the protein and with the surrounding
medium and thus are not strongly damped. 

In this work we propose to model the dynamic
modes by mapping quantitatively an $\alpha-$helix onto a
periodic frame
that supports solitons, like a Toda lattice. Since a Toda lattice
involves only local
interaction and allows one to describe solitons as explicit analytical
solutions, such a model would provide a useful tool, both for quick
numerical simulations and feasible mathematical approaches. Here we
present a numerical study with special emphasis
on the spatio-temporal behavior of the full helix, and we shall neglect explicitly considering
the internal excitations in the (H-N-C=O) units. 

The dynamical behavior of an $\alpha-$helix has been much studied in
the past, particularly using 
simplified 1D models \cite{Davydov73,scott92}.
The main emphasis has been on 
the, so-called Davydov soliton, which is related to the C=O excitation at $\nu
\sim 1650$ cm$^{-1}$.
A recent study using  only non-linear coupling between the C=O and O$\cdots$H excitations
investigated the effects of the 3D coupling \cite{Hennig02}. 
The interaction model is rather different from ours, and in
particular the helix was {\it assumed} to be confined in a narrow cylinder. 

Here we concentrate mainly on the excitations along the hydrogen bonds
- and we have found, among other things, 
the soliton excitations to be phase locked, and hence the excited
helix is spontaneously confined
in a narrow cylinder.

\section{Physical considerations}

\begin{figure}\center

\epsfig{file=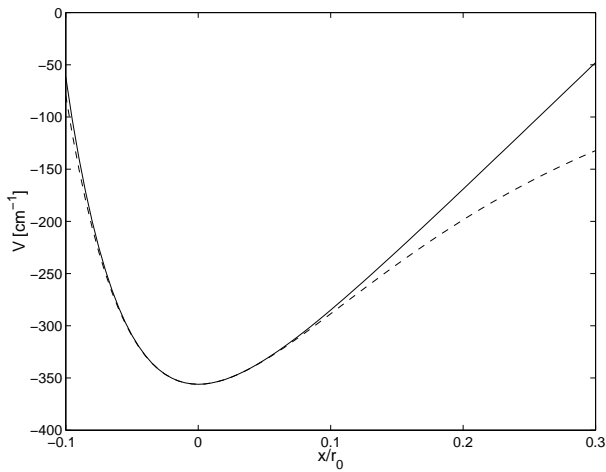,clip=,width=.45\textwidth}
	\caption{Comparison between the Lennard-Jones potential
	(dashed line) and a fit with Toda potential (continuous line)
	for the hydrogen bond. The agreement is good for physically relevant excitation 
	energies $\sim 100$ cm$^{-1}$
	}

	\label{fig:todapot}
\end{figure}
 An essential feature of the helices in a protein is the
         hydrogen bond structure that keeps the helix stable. Of course
         the basic structure of the helix is the poly-peptide backbone
         that is winded up in a homogeneous spiral whose pitch or
         residues per turn determine what type of helix being present,
         be it $\alpha-$helix (the most common type) or a $\pi-$helix.

         These bonds, especially in the case of the
         $\alpha-$helix type where the hydrogen bonds run almost parallel
         to the helix axis, can be regarded as a lattice where the
         interaction between the constituents
         is a typical Lennard-Jones potential describing the
         Van der Walls forces. The interaction can when
         expanded up to the next lowest order including the
         cubic term be mapped onto 
         the studied Toda lattice.

 In one dimension and around an
equilibrium position at $r_{0}$, a Toda potential has the following
form:

\begin{equation}
V(r)=\frac{a}{b}e^{-bx}+ax ,
\end{equation} 

where $x=r-r_0$ is the displacement from the equilibrium and $a$ and $b$ are
two parameters. As we can see from Fig.\ \ref{fig:todapot}, a Toda
potential is asymmetric in a way similar to a Lennard-Jones. 
It does not become flat at large distances but at short ranges it may
be used to model the hard core repulsion on one side of the equilibrium
and the weaker interaction on the other one. An
expansion around the equilibrium $r=r_0$ gives:

\begin{equation}\label{eq:exp}
V(r)=
\frac{a}{b}+\frac{1}{2} abx^2  -\frac{1}{6} ab^2x^3 + o(x^4),
\end{equation} 

showing that the product $ab$ corresponds to the force constant  $k$ in a
harmonic approximation. By equating
the coefficients of the Toda expansion \eq{eq:exp} to the expansion of
a Lennard-Jones potential for the hydrogen bond, $a$ and $b$ can be
estimated yielding $a=k r_0/21$
and $b=21/r_0$.

The harmonic frequency $\nu$ of a
  phonon (at maximum density of states) is given by:

\begin{equation}\label{eq:linmod}
2\pi\nu=2\sqrt{k/m},
\end{equation}

In a chain of amino acids connected by hydrogen bonds O$\cdots $H,
$k\approx1.41\,10^{4}$ dynes/cm and $m=1.7\,10^{-22}$g is the average mass of the
residues \cite{lee98}. 
This estimation gives $\nu=97$ cm$^{-1}$. A complete normal mode
calculation for an infinite poly(L-alanine) $\alpha-$helix gives a
peak exactly at 118 cm$^{-1}$ \cite{lee98}.

A similar fit may
be given for the peptide bond and provides two Toda constants $c$ and $d$
with a corresponding force constant roughly 40 times the one of the
hydrogen bonds \cite{lindgaard02} (simulations with other choices of the parameters for the
peptide bonds have been done, up to a $\pm 20\%$ parameter
change, yielding qualitatively similar results). It is in this range that the
Toda lattice can sustain stable solitons and hence give  an argument for
considering soliton dynamics on the hydrogen bond network.
As we shall in the following sections, the two bonds have a very different
role on the energy propagation. The hydrogen bonds provide three one dimensional,
nonlinear lattices, where solitons appear, while the peptide bonds act as a strong coupling
among the three lattices.

\section{Model description}

\begin{figure}\center

\epsfig{file=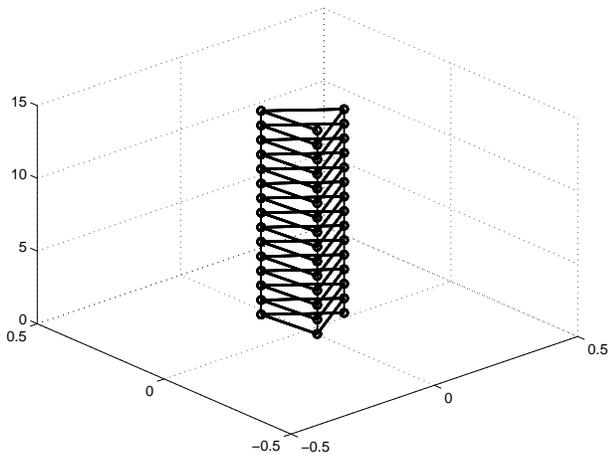,clip=,width=.45\textwidth}
	\caption{The system modeled. Each dot is an
	amino acid, {\it i.e.} the (H-N-C=O) unit including various side chains bound to the C-atom.
	The heavy and thin lines represent respectively
	the peptide and hydrogen bonds. The space unit has been
	normalized to the equilibrium distance of the hydrogen
	bond. The scale has been enlarged on 
	the $x-y$ plane. The peptide bonds connect all the amino acids
	in a spiral. The hydrogen bonds connect the amino acids in
	three parallel chains.}
	\label{fig:alpha}
\end{figure}

In this work we thus modeled an $\alpha-$helix with the aim of
investigating the propagation of phonons and solitons along the hydrogen bond
and the effect of the coupling with the peptide bond.
Direct integration of the
equation of motion have been made. A picture of the system modeled is
shown in Fig.\ \ref{fig:alpha}. The Hamiltonian of the system is
given by adding together the kinetic energy, the potential energy of the
three chains with the hydrogen bonds and the potential energy of the
peptide bonds. Calling $x_j$ and $p_j$ the space
coordinate and the momentum of the $j$-th amino acid and numbering the
amino acids as
they appear along the spiral, the Hamiltonian is given
by:

\begin{eqnarray}\label{eq:hprot}
H=E_{\text{kin}}+
V_{\text{H}}+V_{\text{peptide}}=\nonumber\\
=\frac{1}{2}\sum_{j=1}^N p_j^2+
\sum_{j=1}^{N}V_{a,b}(x_j,x_{j+1},x_{j-1})\nonumber\\
+\sum_{j=1}^{N}V_{c,d}(x_j,x_{j+3},x_{j-3}),
\end{eqnarray}

where $V$ are Toda potentials of parameters $a,b$ and $c,d$ and 
have obviously to include only the amino acids with
$j-1>0,\quad j-3>0,\quad j+1<N,\quad j+3<N$. The equation of
motion are obtained straightforwardly.

\section{Solitons along an the H-bonds}
\begin{figure}
\center
\epsfig{file=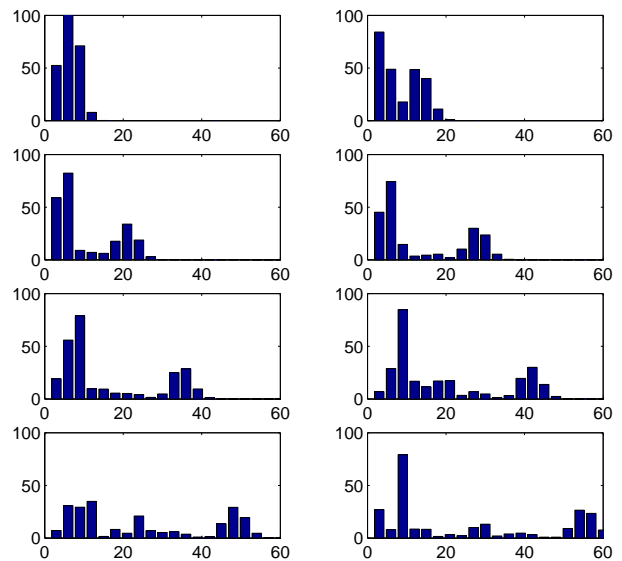,clip=,width=.45\textwidth}
	\caption{Propagation of a perturbation along the hydrogen
	bonds, in presence of the peptide bonds. Although a large
	amount of energy remains localized
	close to the perturbed amino acid, a soliton is formed spontaneously. The
	perturbation has an energy of $618$ cm$^{-1}$. The
	snapshots are taken every 0.1 picoseconds. The other two
	hydrogen chains not shown have an analogous behavior.
	\label{fig:barh1}}
\end{figure}

\begin{figure}
\center
\epsfig{file=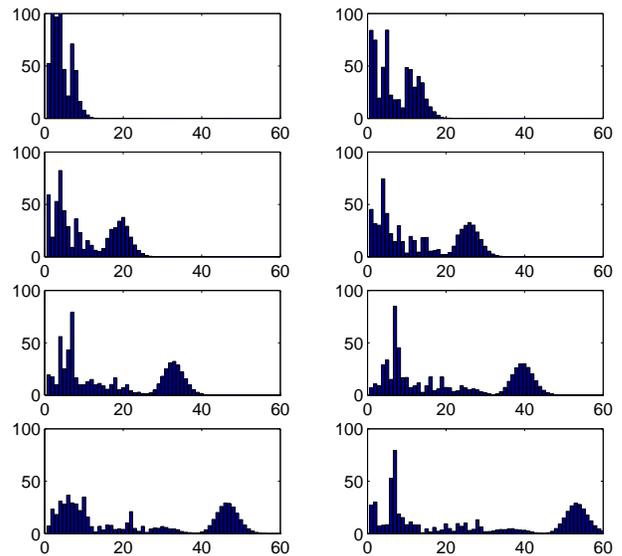,clip=,width=.45\textwidth}
	\caption{Energy flow along the whole helix. The three solitons
	along the hydrogen bonds compose a united soliton, traveling
	along the whole system. 
	\label{fig:barcov}}
\end{figure}

\begin{figure}
\center
\epsfig{file=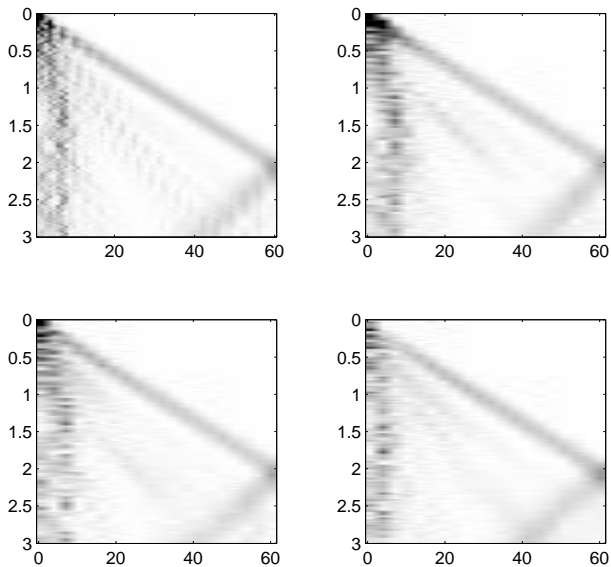,clip=,width=.45\textwidth}
	\caption{Space-time plot of the energy flow. The top-left picture
	gives all the amino acids, while the others show the amino acids
	belonging to the same chain of hydrogen bonds. Time is in picoseconds.
	\label{fig:solcovst}}
\end{figure}

Let us consider an $\alpha$-helix with a perturbation along one of H-bond chain. The
energy flow along the helix is shown in
Figs. \ref{fig:barh1}-\ref{fig:barcov}. Due to the presence of the
peptide bonds, the
perturbed chain is coupled with the other two
chains present in the helix. As a consequence, not all the energy
travels along the chain in a localized way, but part of it remains
close to the perturbation point and spreads into the system at a
slower speed.
Another interesting phenomenon resulting from the fact that the three
chains are bounded can be seen looking at the energy flow along the
other two chains (Fig. \ref{fig:solcovst}). We see that some
energy is soon transferred along the peptide bond to other chain. As
in the case of the perturbed chain, we can identify two waves: one
fast and localized and the other slow. The three solitons along the hydrogen
bonds are traveling together. If the energy flow of the entire helix
is plotted (Fig. \ref{fig:barcov}), one finds that the solitons of the
three hydrogen chains compose a united triple-soliton, traveling along the
axis of the whole
helix.

The basic mechanism of the triple soliton solution is given by the two
different roles of the hydrogen and peptide bonds. The hydrogen bond
provides three one-dimensional lattices that can support solitary
waves. On the other hand, the peptide bond acts as a coupling among
the H-bond chains: it induces solitons from one chain to another and
entrains them, but otherwise does not qualitatively affect their dynamics. This
observation can be verified noticing that, after the triple wave is formed, each soliton behaves as it
was on an independent, one dimensional lattice with the constant of
the hydrogen bonds. In fact, the dynamics of a soliton on a one
dimensional Toda lattice is characterized  by the following relations
\cite{toda81}.
The energy is:

\begin{equation}
E=\frac{2a}{b}(\sinh{\kappa}\cosh{\kappa}-\kappa);
\end{equation}

the profile, in terms of the displacements $|x_j|$ from the equilibrium distance is given by:

\begin{equation}
e^{-b x_j}-1=\frac{m}{ab}\beta^2\sech^2(\kappa j\pm \beta t);
\end{equation}

finally the speed $v$ is:

\begin{equation}
v=\sqrt{\frac{a b}{m}}\frac{\sinh \kappa}{\kappa}.
\end{equation}

In such relations, $\beta=\sqrt{ab/m}\sinh{\kappa}$ and $\kappa$ is a parameter
that completely characterize the soliton dynamics and shape
($\kappa$ being proportional to the width of the soliton). If
$\kappa$ is computed by fitting the energy, the profile, and
velocity of one of the soliton \emph{using the parameter of the H-bond
chain only}, approximately the same value is obtained in all the three
fitting: respectively, 0.74,
0.85, and 0.78. Using the latter, this correspond to an energy of $E=41$ cm$^{-1}$ 
distributed over about 8 sites (see fig.\ref{fig:barcov}) and a velocity of 
$v=1.10\; v_s$, where the sound velocity $v_s= 1.73\; 10^5$
cm/sec.

In a perfect (infinite, 1D) Toda lattice solitons and periodic
waves (sinusoidal in the limit of small energies)
can exist simultaneously, with infinite life time \cite{toda81}. Small deviations from
the ideal picture will lead to 
a small coupling and hence exchange of energy between the two modes.
Moreover, the periodic wave can be obtained as a superposition of solitons \cite{toda81}. 
The dispersion relation between the
wavelength $\lambda$ and the frequency $\nu$ of the periodic wave is given
by:

\begin{equation}
2 F \nu=\sqrt{\frac{ab}{m}}\bigg/\sqrt{\frac{1}{\sn^2(2 F \lambda)-1}+\frac{G}{F}},
\end{equation}

where $\sn$ is a Jacobian elliptic function and $F$ and $G$ are two
parameters that depends on the profile.
At the found relevant energies and deviations the Toda and the Van der
Walls potential 
are almost identical, see Fig. 1. Hence we expect the obtained results
to be essentially valid also 
for the latter more realistic potential. 

Other simulations, performed by initializing the system with
 different energies, show solitons of different width, but again in agreement
with the 1D Toda model. 
We have also studied  the
effect of changing the parameters of the hydrogen and
peptide bonds and found that only the hydrogen bond
parameters effects the dynamics and shape of the solitons.

We conclude that the peptide bond is important only for
creating and entraining the three solitons, while the behavior of the coupled
solitons agree to a good
approximation with respect to the dynamics and shape to that one would
have on the uncoupled, 
1D lattice of the H-bonds. In particular, we do not see a concentration of energy on
one strand, as reported by Hennig 
\cite{Hennig02} for strong non-linear coupling. In fact the opposite:
An excitation applied to one strand
results in phase-locked solitons moving in parallel on all three
strands. That is important
for keeping the $\alpha$-helix confined in space. For some
perturbation we have also observed a train of solitons, emitted
periodically by the distortion mode.

\begin{figure}
\center
\epsfig{file=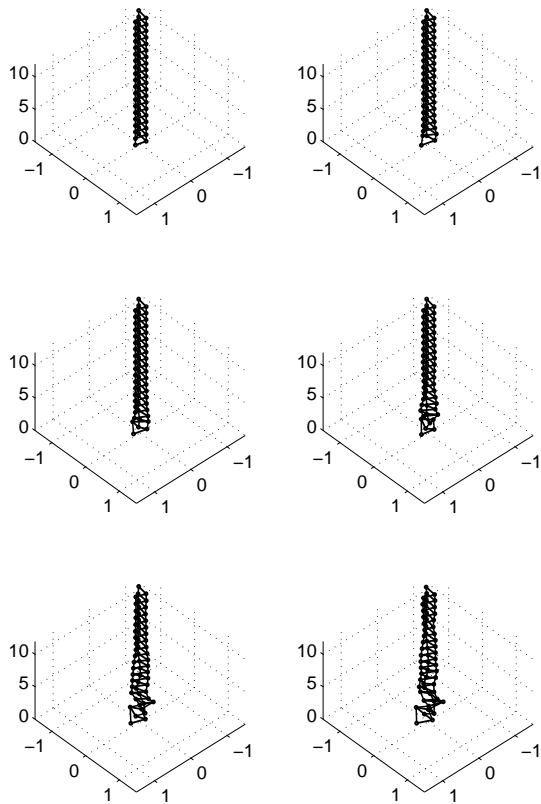,clip=,width=.4\textwidth}
	\caption{Conformational change of the $\alpha-$helix after a
	perturbation that gives an impulse to the first amino acid
	(along the axial direction and toward the helix). There are
	two responses: a large and slow 
	distortion mode, and a localized and quickly (supersonically)
	moving pulse, due to the soliton
	(for the latter see the enlargement in Fig.\
	\ref{fig:sol6z}). A perturbation given in the opposite direction
	also gives rise to a qualitatively similar phenomenon.
	Snapshots at t=0, 0.3, 0.6, 1.2, 2.4 and 3 picoseconds.
	See Fig.\
	\ref{fig:solcovst} for an overview of the energy flow along
	the helix.
	\label{fig:sol6}}
\end{figure}
\begin{figure}
\center
\epsfig{file=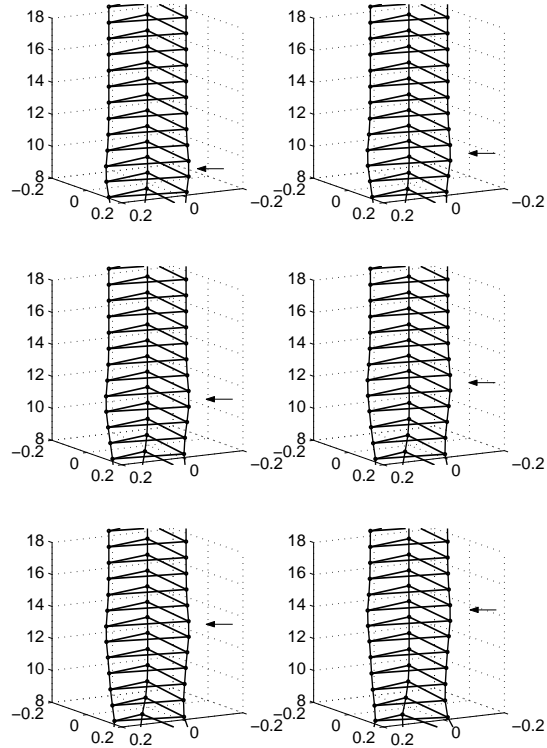,clip=,width=.4\textwidth}
	\caption{Enlargement of Fig.\ \ref{fig:sol6}. The localized
	wave corresponding to the triple soliton traveling on the hydrogen
	chains along the helix. Snapshots every 0.6 picoseconds starting from t=0.8 picoseconds.
	The arrows show the position of the soliton. 
	\label{fig:sol6z}}
\end{figure}

\section{Discussion}

An $\alpha-$helix can be modeled by three coupled Toda lattices.
The specific topology of this system gives rise to
peculiar collective oscillations that are of great relevance for
proteins, since they control their structure formation and may be
also related to their folding/unfolding behavior. In this work we have focused on a mode that has been
observed in the pump-probe experiments and, given its relatively low energy $\sim 100$ cm$^{-1}$, has been
related to
extended, collective modes. An unexplained feature of this mode is its
long life. This is surprising, since a collective mode, involving the 
motion of a lot of amino acids should have a strong interaction with
the rest of the protein and the solvent, and thus decay quickly.
Following the suggestion in \cite{lindgaard02},
this problem has
been approached proposing that the behavior of such a mode involves considering also soliton solutions,
{\it i.e.}, localized waves that travel along the hydrogen bond chains of
the helix and hence has a
small interaction with the surrounding. We observed
two types of waves. The soliton on one hydrogen chain induces a soliton on each
of the other two hydrogen chains to which is coupled through the
peptide bond. The three solitons propagate together in a single,
localized, and fast wave along the helix. A second type of wave also appear in
the system, as a comparatively slower distortion mode.
Let us now discuss the two mechanisms in connection with the
suggestion of \cite{lindgaard02}. It is useful to look at the
conformational change corresponding to the two waves
(Figs. \ref{fig:sol6} and \ref{fig:sol6z}). The solitary wave is then
especially interesting. In fact, while the distortion mode
results in a large conformational change and for this reason may be quickly
damped by the interaction with the surrounding, the solitary wave allows to
keep an amount of energy over the helix with a minimal conformational
change. The solitons in the actual $\alpha$-helix are not perfect and will slowly
disperse energy into excitations involving motion of the same
atoms. That is in particular to 
the phonons exciting the very same units in an oscillatory motion
along the H-strands with energies up 
to $\sim 100$ cm$^{-1}$. This fact,
with the observation that
the three locked solitons appears spontaneously, gives support to the
idea proposed in \cite{lindgaard02}.

Although we have not explicitly considered the internal excitations in
the (H-N-C=O) units, it is clear
that the energy may also be fed into the C=O excitation at $\nu \sim
1650$ cm$^{-1}$; the soliton mechanism here described hence may also 
provide a means for having unexpected long life time of that
mode. This fact has been observed
and is reported in this volume by Austin {\it et al.} and in Ref. \cite{xie00}.
More work is in progress to elaborate on the presented model.

\bibliography{artbib}

\end{document}